\documentclass[pra,nofootinbib,floats,aps,twocolumn, tightenlines,superscriptaddress,longbibliography]{revtex4-2}
\usepackage[utf8]{inputenc}
\usepackage{amsmath}
\usepackage{amsfonts}
\usepackage{dsfont} 
\usepackage{graphicx}
\usepackage{todonotes}
\usepackage{gensymb}
\usepackage{bbold}
\usepackage{bm}
\usepackage{natbib}
\usepackage{amssymb}
\usepackage{latexsym}
\usepackage{fancyhdr}
\usepackage[bbgreekl]{mathbbol}
\usepackage{epsfig}
\usepackage{color}
\usepackage{textcomp}
\usepackage{amsthm}
\usepackage{comment}
\usepackage{ulem}
\usepackage{hyperref}
\usepackage[nice]{nicefrac}

\usepackage{mathrsfs}

\graphicspath{{Pictures/}}

\usepackage{subfigure}

\newcommand{\be}{\begin{equation}}
\newcommand{\ee}{\end{equation}}

\newcommand{\stkout}[1]{\ifmmode\text{\sout{\ensuremath{#1}}}\else\sout{#1}\fi}

\begin{document}

\title{Experimental device-independent quantum key distribution between distant users}

\author{Wei Zhang}
\altaffiliation{These authors contributed equally}
\affiliation{Fakult{\"a}t f{\"u}r Physik, Ludwig-Maximilians-Universit{\"a}t, M{\"u}nchen, Germany}
\affiliation{Munich Center for Quantum Science and Technology (MCQST), M{\"u}nchen, Germany}

\author{Tim van Leent}
\altaffiliation{These authors contributed equally}
\affiliation{Fakult{\"a}t f{\"u}r Physik, Ludwig-Maximilians-Universit{\"a}t, M{\"u}nchen, Germany}
\affiliation{Munich Center for Quantum Science and Technology (MCQST), M{\"u}nchen, Germany}

\author{Kai Redeker}
\altaffiliation{These authors contributed equally}
\affiliation{Fakult{\"a}t f{\"u}r Physik, Ludwig-Maximilians-Universit{\"a}t, M{\"u}nchen, Germany}
\affiliation{Munich Center for Quantum Science and Technology (MCQST), M{\"u}nchen, Germany}

\author{Robert Garthoff}
\altaffiliation{These authors contributed equally}
\affiliation{Fakult{\"a}t f{\"u}r Physik, Ludwig-Maximilians-Universit{\"a}t, M{\"u}nchen, Germany}
\affiliation{Munich Center for Quantum Science and Technology (MCQST), M{\"u}nchen, Germany}

\author{Ren{\'e} Schwonnek}
\affiliation{Department of Electrical \& Computer Engineering, National University of Singapore, Singapore}
\affiliation{Naturwissenschaftlich-Technische Fakult{\"a}t, Universit{\"a}t Siegen, Germany}

\author{Florian Fertig}
\affiliation{Fakult{\"a}t f{\"u}r Physik, Ludwig-Maximilians-Universit{\"a}t, M{\"u}nchen, Germany}
\affiliation{Munich Center for Quantum Science and Technology (MCQST), M{\"u}nchen, Germany}

\author{Sebastian Eppelt}
\affiliation{Fakult{\"a}t f{\"u}r Physik, Ludwig-Maximilians-Universit{\"a}t, M{\"u}nchen, Germany}
\affiliation{Munich Center for Quantum Science and Technology (MCQST), M{\"u}nchen, Germany}

\author{Valerio Scarani}
\affiliation{Centre for Quantum Technologies, National University of Singapore, Singapore}
\affiliation{Department of Physics, National University of Singapore, Singapore}

\author{Charles C.-W. Lim}
\altaffiliation{charles.lim@nus.edu.sg}
\affiliation{Department of Electrical \& Computer Engineering, National University of Singapore, Singapore}
\affiliation{Centre for Quantum Technologies, National University of Singapore, Singapore}

\author{Harald Weinfurter}
\altaffiliation{h.w@lmu.de}
\affiliation{Fakult{\"a}t f{\"u}r Physik, Ludwig-Maximilians-Universit{\"a}t, M{\"u}nchen, Germany}
\affiliation{Munich Center for Quantum Science and Technology (MCQST), M{\"u}nchen, Germany}
\affiliation{Max-Planck Institut f{\"u}r Quantenoptik, Garching, Germany}

\date{\today}

\begin{abstract}
\noindent Device-independent quantum key distribution (DIQKD) is the art of using untrusted devices to establish secret keys over an untrusted channel. So far, the real-world implementation of DIQKD remains a major challenge, as it requires the demonstration of a loophole-free Bell test across two remote locations with very high quality entanglement to ensure secure key exchange. Here, we demonstrate for the first time the distribution of a secure key---based on asymptotic security estimates---in a fully device-independent way between two users separated by 400 metres. The experiment is based on heralded entanglement between two independently trapped single Rubidium 87 atoms. The implementation of a robust DIQKD protocol indicates an expected secret key rate of $r=0.07$ per entanglement generation event and $r>0$ with a probability error of $3\%$. Furthermore, we analyse the experiment's capability to distribute a secret key with finite-size security against collective attacks.
\end{abstract}

\maketitle

\section{Introduction}
Quantum key distribution (QKD)~\cite{Bennett2014,Ekert1991} uses the unique features of quantum mechanics to exchange provably-secure secret keys over an untrusted network. The technology is now well established in a wide variety of network settings~\cite{Elliot2007,Peev2009,Sasaki2011,Zhang2018,Boaron2018,Schmitt-Manderbach2007,Ursin2007,Nauerth2013,Liao2017,Liao2018} and commercial QKD systems are available as well. QKD protocols are designed to detect eavesdropping attacks on the quantum channel through intrinsic quantum effects like the no-cloning theorem and the uncertainty principle. In order to invoke these effects, most QKD protocols require that the underlying quantum devices are well characterised and are accurately described by the mathematical models used in the security analysis. The users of such protocols must trust not only the honesty of the QKD vendors, but also the specifications provided by them. This may be critical: indeed, it has been known for at least a decade that some QKD devices can be readily hacked from the outside by exploiting physical features that had not been deemed relevant in a first analysis~\cite{XuRMP2020}.

To overcome the above issues, a promising solution is to use device-independent QKD (DIQKD)---a correlation-based method which allows the users to exchange secret keys with uncharacterised (or untrustworthy) quantum devices. First introduced by Mayers and Yao~\cite{mayers1998quantum}, DIQKD~\cite{acin2007device,Pironio2009,Barrett2005,reichardt2013classical,Vazirani2014,miller2016robust,Arnon-Friedman2018} ensures the proper and secure functioning of the underlying QKD devices via a loophole-free Bell test~\cite{Bell1965, Brunner2014RMP}. More specifically, the users only need to analyse their input-output measurement statistics to put an upper bound on the amount of information leaked to an eavesdropper, hence eliminating the requirement to characterise the devices. 
Thus, DIQKD automatically provides security against implementation flaws and especially any form of misalignment. It only requires a few very basic assumptions to be fulfilled. 

\begin{figure}
\centering
\includegraphics[width=0.45\textwidth]{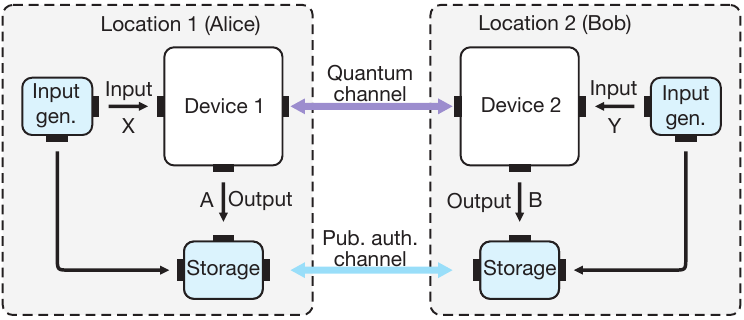}
\caption{\textbf{Schematic of the DIQKD connection.} Each of the two parties, Alice and Bob, holds one of the two QKD devices which are connected via a quantum channel. The devices receive the inputs $X$ and $Y$, and respond with outputs $A$ and $B$, respectively. To run the protocol each party needs a trusted supply of inputs and a trusted local storage unit to store both output and inputs. Additionally, a trusted authenticated public channel between the two parties is necessary for exchange of information during post-processing.}
\label{fig:DI-scheme}
\end{figure}

For ease of reference, let us list these basic requirements of DIQKD (see the Supplemental Material A for more details). The two DIQKD users, Alice and Bob, (i) should each hold a device that is able to receive an input and then respond with an unambiguous output, as illustrated in Fig.~\ref{fig:DI-scheme}. The communication between these devices is assumed to be restricted, namely (ii) the users control when their respective devices communicate with each other~\cite{Arnon-Friedman2019}; and (iii) the devices do not send unauthorised classical information to an eavesdropper. Finally, as it is with any QKD protocol, it is required that (iv-a) quantum mechanics is correct, (iv-b) the users' inputs are private and random, and (iv-c) the users are connected via an authenticated classical channel and the post-processing platform is trusted.

The experimental realisation of DIQKD is, however, a major challenge. The main difficulty is to devise a system that enables for a loophole-free Bell test while achieving both an high Bell violation and a low quantum bit error rate (QBER). Current state-of-the-art loophole-free Bell experiments~\cite{Hensen2015,Giustina2015,Shalm2015,Rosenfeld2017} are able to achieve significant Bell violations, but the QBERs are still not good enough for DIQKD (e.g., see the survey provided by Ref.~\cite{Murta2019}). To lower these requirements, one approach is to devise DIQKD protocols which are more robust and efficient. Recently, this yielded two improved variants of the original DIQKD protocol---one based on noisy-preprocessing~\cite{ho2020noisy} and the other based on randomised key setting~\cite{Schwonnek2020}.

Here we present a proof-of-concept DIQKD experiment with distant users, demonstrating the protocol proposed in Ref.~\cite{Schwonnek2020}. For this, we employ the upgraded version of an event-ready loophole-free Bell experiment \cite{Rosenfeld2017}. Here, the quantum channel for DIQKD is formed by two single $^{87}$Rb atoms separated 400 metres geographically (the two laboratories are connected via a 700 m long optical fibre). The event-ready entanglement generation scheme runs in two stages for each measurement round: (1) an entangled spin-polarisation state (between the atom and a spontaneously emitted photon) is first generated locally in each laboratory and (2) then the photons are sent to a Bell-state measurement setup for entanglement swapping. Hence, whenever the entanglement swapping is successful, entanglement between the spin states of the atoms is generated and announced. Key improvements in the entanglement generation rate, coherence of atomic states, and entanglement swapping fidelity enabled the implementation of the protocol. Based on the measurement data obtained from the experiment, we find that a positive asymptotic secret key fraction (the ratio of achievable secret key length to the total number of heralded events) of 0.07 was achieved in a fully device independent configuration.

\begin{figure*}
\centering
\subfigure{\includegraphics[width=0.67\textwidth]{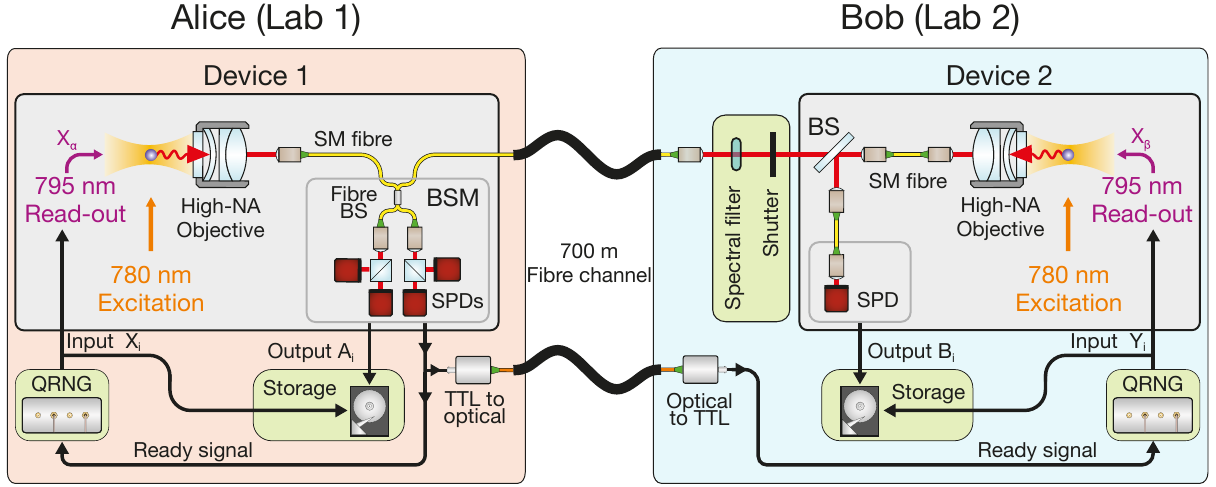}} \quad
\subfigure{\includegraphics[width=0.3\textwidth]{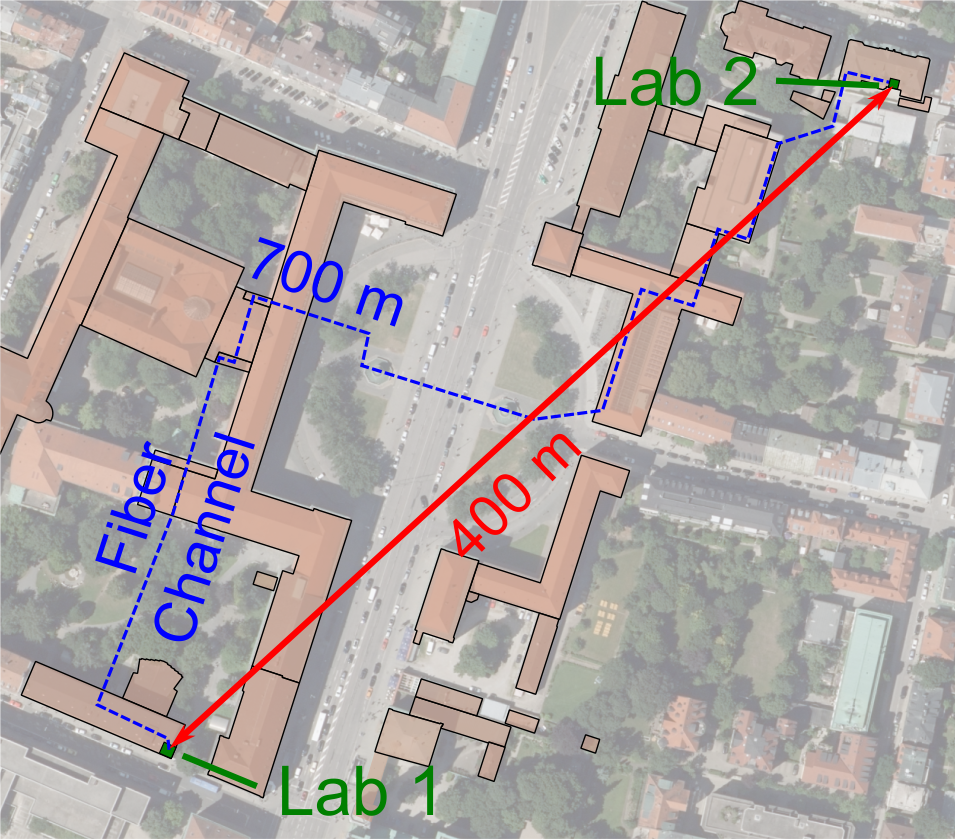}}
\caption{\textbf{Overview of the quantum network link.}
(\textbf{a}) Device~1 in Lab~1 (Alice) is formed by a single atom trap setup and a Bell-state measurement (BSM) setup. Device~2 in Lab~2 (Bob), located 400 metres away, is formed by a second single atom trap setup together with 90:10 (T:R) beamsplitter (BS) and a single photon detector (SPD). Each trap setup contains a high-NA objective to optically trap a single atom and collect atomic fluorescence which is coupled into a single-mode (SM) fibre. The atoms are entangled by synchronously exciting them after which the spontaneously emitted photons are guided to the BSM to employ an entanglement swapping protocol. The BSM is formed by a 50:50 fibre BS to spatially overlap the photons, two polarising BSs, and four SPDs. 
Coincidental detection of a single photon on two detectors in the same output arm of the fibre BS heralds the state $|\Psi^+\rangle$, which is communicated to both parties via a ready signal. After receiving the ready signal, two quantum random number generators (QRNG)~\cite{Fuerst2010} select the inputs to the devices which determines the polarisation of a $795~\text{nm}$ readout pulse in a by polarisation controlled state selective ionisation scheme. Whether or not the ionisation try was successful yields the binary output of the devices which is, since ionised atoms are lost from the trap, determined by fluorescence collection. The in- and outputs of each round are stored locally using a trusted storage. A spectral filter and shutter are implemented in Lab 2 to avoid leakage of information on the setting and the measurement result via the quantum channel. (\textbf{b}) Map showing the main campus of the LMU in Munich indicating the locations of the two used laboratories. Map data were provided by Ref. \cite{mapLMU_ref}.}
\label{fig:ExpScheme}
\end{figure*}

\section{Methods}

\subsection{DIQKD protocol} The protocol considered here is similar to the standard protocol~\cite{Acin2007,Pironio2009}, except that two measurement settings are used for key generation instead of one. Importantly, in doing so, the protocol can tolerate significantly more noise---the critical QBER is extended from 7.1\% to 8.2\%~\cite{Schwonnek2020}. The protocol considers that Alice and Bob each hold a device, which are connected via a quantum channel (Fig.~\ref{fig:DI-scheme}). The protocol consists of $N$ measurement rounds, whereby in each $i$th round both devices receive an input ($X_i$ and $Y_i$) and respond with an output ($A_i$ and $B_i$). More specifically, Alice's device accepts four different values $X_i\in\{0,1,2,3\}$, while Bob's device has two inputs $Y_i\in\{0,1\}$. The input for each round is provided by a trusted local source of randomness. Both devices output two possible values, $A\in\{\uparrow,\downarrow\}$ at Alice's side and $B\in\{\uparrow,\downarrow\}$ at Bob's side. Additionally, the input and output values are recorded and stored into a local secured storage. 

After $N$ rounds the users stop the measurements and begin with classical post-processing. For this, Alice and Bob reveal their inputs for each round over an authenticated public channel. For the rounds with differing input settings, i.e., $X\in\{2,3\}$ together with $Y\in\{0,1\}$, the outputs are shared over the public channel to compute the Clauser-Horne-Shimony-Holt (CHSH)~\cite{Clauser1969} value using
\begin{equation}
    S:=E_{2,1}-E_{2,0}-E_{3,0}-E_{3,1},
\label{eq:S}
\end{equation}where the correlation functions are defined as $E_{X,Y}:=\big(N^{A=B}_{X,Y}-N^{A\neq B}_{X,Y}\big)/N_{X,Y}$. Here, $N_{X,Y}$ is the number of rounds with input combination $X,Y$, while $N^{A=B}_{X,Y}$ is the number of rounds with identical outcomes and input combination $X,Y$. Provided that the devices share a sufficiently entangled state, the Bell inequality can be violated, i.e., $S>2$. In our experiment, we target the generation of the maximally-entangled state
\begin{equation}
|\Psi^+\rangle_{AB}=\frac{|\uparrow\rangle_{z,A}|\downarrow\rangle_{z,B}+|\downarrow\rangle_{z,A}|\uparrow\rangle_{z,B}}{\sqrt{2}},
\label{eq:atomatom}
\end{equation} where the orthogonal spin states $|\uparrow\rangle_{z}$ and $|\downarrow\rangle_{z}$ are defined as the computational basis states of the protocol (see next section for more details).

The raw data are sifted so that only the outputs of measurement rounds with identical input settings are kept for further processing. The QBERs for both key settings are denoted by $Q_{0}=N^{A=B}_{0,0}/N_{0,0}$ for $X_i=Y_i=0$ and $Q_{1}=N^{A=B}_{1,1}/N_{1,1}$ for $X_i=Y_i=1$. Note that the key pairs are supposed to be anti-correlated due to the use of anti-correlated entangled states. Both the QBERs $(Q_0,Q_1)$ and the CHSH value $S$ are used to determine the amount of information about the sifted key that could have been obtained by an eavesdropper ~\cite{renner2008security}. Next, by applying a technique known as leftover hashing, the eavesdroppers (quantum) information about the final key can be reduced to an arbitrary low level, defined by the security error of the protocol~\cite{tan2020improved}. In this experiment, we focus on estimating the asymptotic security performance of the considered DIQKD protocol. For this purpose, we note that in the asymptotic limit and in case of a depolarising quantum channel, positive key rates can be achieved when the expected CHSH value satisfies $S>2.362$ (or equivalently, $Q<0.082$ with $Q_0=Q_1=Q$) as shown in Ref~\cite{Schwonnek2020}.

\subsection{Quantum network link} A quantum network link (QNL) generates the entanglement between the two spatially separated laboratories in order to implement the DIQKD protocol. As mentioned before, in our setup, entanglement is generated between two optically trapped single $^{87}$Rb atoms located in laboratories 400 m apart and connected via a 700 metre long optical fibre channel, see Fig.~\ref{fig:ExpScheme}. The atoms act as quantum memories where a qubit is encoded in the Zeeman-substates of the $5\text{S}_{1/2}|\text{F}=1,m_\text{F}=\pm1\rangle$ ground state, with $m_\text{F}=+1$ and $m_\text{F}=-1$ designated as computational basis states, $|\uparrow\rangle_z$ and $|\downarrow\rangle_z$, respectively, and where the quantization axis $\hat{z}$ is defined by the fluorescence collection setup.

\begin{figure}
\centering
\protect\includegraphics[width=0.48\textwidth]{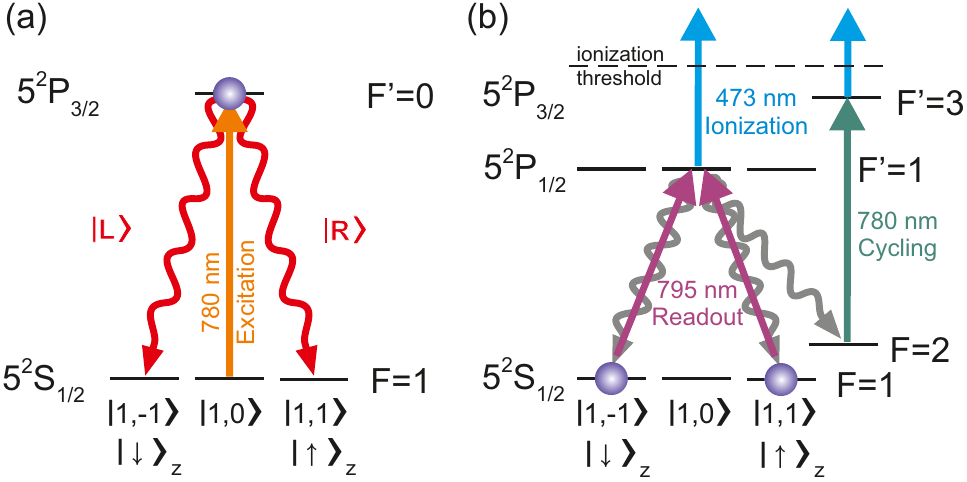}
\caption{\textbf{Schematics of the entanglement generation and atomic state readout schemes.} \textbf{a}, An entangled atom-photon state is generated by the spontaneous decay subsequent to excitation to the state $5^2\text{P}_{3/2}|\text{F}'=0,m_{\text{F}'}=0\rangle$. \textbf{b}, The qubit state is read-out via a state dependent ionisation scheme. First, a by polarisation controlled superposition of the qubit state is excited to the $5^2\text{P}_{1/2}$ level by a $795~\text{nm}$ laser pulse. Subsequently, the excited atom is ionised by a second laser pulse with a wavelength of $473~\text{nm}$. If the atom decays to the state $5^2\text{S}_{1/2}|\text{F}=2\rangle$ before it is ionised, it is excited to the state $5^2\text{P}_{3/2}|\text{F}'=3\rangle$, which is ionised as well.}
\label{fig:schemes}
\end{figure}

% atom-atom generation: inital state preparation, atom-photon, atom-atom
The two distant atoms are entangled using an entanglement swapping protocol~\cite{Hofmann2012}. The sequence starts by synchronously exciting a single atom in each trap to the state $5^2\text{P}_{3/2}|\text{F}'=0,m_{\text{F}'}=0\rangle$; when decaying back to the ground state, each of the atomic qubits becomes entangled with the polarisation of the respective spontaneously emitted single photon (Fig.~\ref{fig:schemes}a). This results in an entangled atom-photon state $|\Psi\rangle_{AP} = 1/\sqrt{2}(|\downarrow\rangle_x|V\rangle + |\uparrow\rangle_x|H\rangle)$~\cite{Volz2006}, where $|\uparrow\rangle_x:=(|\uparrow\rangle_z + |\downarrow\rangle_z)/\sqrt{2}$ (resp. $|\downarrow\rangle_x:=(|\uparrow\rangle_z - |\downarrow\rangle_z)/\sqrt{2}$), and $|H\rangle$ and $|V\rangle$ mean parallel and orthogonal linear polarisations with respect to the optical table, respectively. The two photons are then guided to a Bell-state measurement (BSM) setup. Projection of the photons onto a $|\Psi^+\rangle$ state heralds the creation of the maximally entangled atom-atom state, as given in equation (\ref{eq:atomatom}). More specifically, given a successful projection, a ready signal is sent to the trap setups and the atomic qubits are measured only after receiving this signal.

The two atomic qubits are independently analysed via a state-selective ionisation scheme
(Fig.~\ref{fig:schemes}b)~\cite{Leent2020}. There, a particular state of the atomic qubit is ionised depending on the polarisation $\chi=\cos(\gamma)V+e^{-i\phi}\sin(\gamma)H$ of a read-out laser pulse ($\gamma = \alpha$ for Alice's and $\gamma = \beta$ for Bob's device) and leaves the trap. If the atom is still in the trap, it is projected onto the state 
\begin{equation}
|D\rangle=e^{i\phi}\cos(\gamma)|\uparrow\rangle_x-\sin(\gamma)|\downarrow\rangle_x.
\label{eq:readout}
\end{equation}
The presence of the atom is then tested using fluorescence collection, which yields the final measurement outcome. On Alice's side, the single-photon detectors (SPDs) of the BSM detect the fluorescence of the atom, while on Bob's side an unbalanced beam splitter directs a small fraction of the florescence light onto a single SPD (Fig.~\ref{fig:ExpScheme}). As such, with this scheme, the detection efficiencies of Alice's and Bob's measurements are effectively one since this final test is performed for every round, any component loss is reflected as noise in the quantum channel.

While the requirements for a DIQKD implementation are less stringent with the newly proposed protocols, significant improvements over existing loophole-free Bell experiments are still required. To that end, we enhance the entanglement generation rate, coherence of atomic states, and entanglement swapping fidelity of the loophole-free setup reported in Ref.~\cite{Rosenfeld2017}. 

Concerning the entanglement generation rate, custom-design high numerical-aperture objectives are installed in each trap to increase the single photon collection efficiency by a factor greater than $2.5$. This ultimately leads to an atom-atom entanglement generation efficiency of $0.49\times10^{-6}$ following an excitation pulse pair. Together with a duty cycle of approximately 1/2 and a repetition rate of the entanglement generation tries of $52$ kHz resulting in an event rate of 1/82 s$^{-1}$. Note that for event-ready entanglement generation schemes the repetition rate of the experiment is limited by the communication times between the two devices and the BSM. For DIQKD protocols, this results in a trade-off between the maximum separation of the users and the achieved secret key rate---not considering multiplexing techniques, see Outlook. 

\begin{figure*}
\centering
\protect \includegraphics[width=0.95\textwidth]{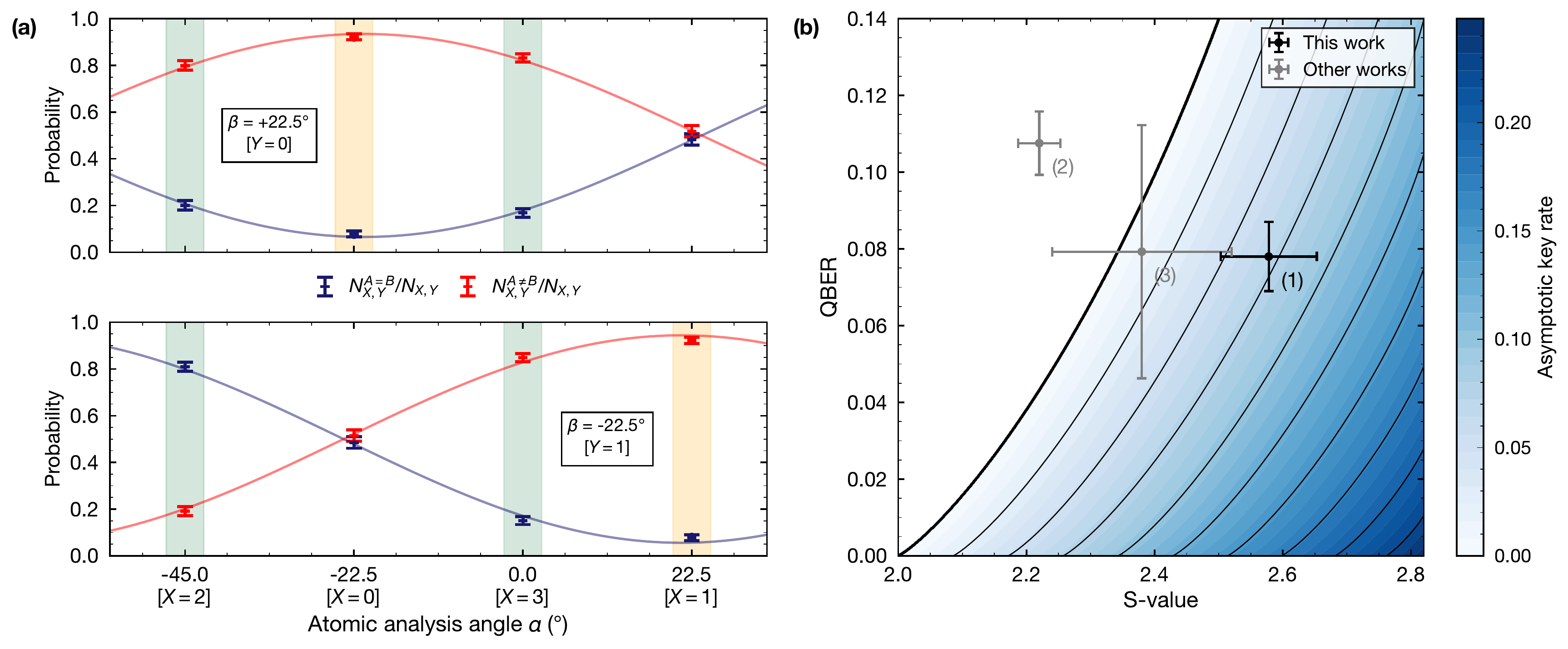}
\caption{\textbf{Observed atom-atom state correlations and the expected secret key rate.} (a) The (anti-)correlation probabilities of the device outputs for the eight input combinations, resulting in $S=2.578(75)$ and $Q=0.0779(91)$. The data is fitted with sinusoidal functions estimating visibilities of $0.869(25)$ and $0.888(45)$. The settings with the green (yellow) background contribute to the evaluation of the $S$ ($Q$). (b) Expected secret key rate for varying CHSH S-value and QBER for the robust DIKQD protocol~\protect\cite{Schwonnek2020}. The presented work (1) shows an expected secret key rate of 0.07 and lies well inside the positive region. For comparison, the results of~\protect\cite{Rosenfeld2017} (2) and~\protect\cite{Hensen2015,Hensen2016} (3), which to the best of our knowledge, are the only experiments that fulfill the requirements for DIQKD with significant distance between the two users. Note that (2) does not reach the positive key regime due to the high QBER. The error bars indicate statistical errors of one standard deviation.} 
\label{fig:result}
\end{figure*}

The coherence and stability of the atomic qubit states are limited by the fluctuations of local magnetic fields and position-dependent vector light shifts which are introduced by the tight focus of the optical dipole traps. The latter is especially crucial as it allows a high fidelity state measurement only when the atom has completed a full oscillation in the trap~\cite{PhD-Daniel}. Here, the better optical components of the new collection setup, which is also used to focus the trapping laser, enable a more symmetric trapping potential. Then, in combination with reduced electrical noise in the active magnetic field stabilization, lowering of atom temperatures, and employing a magnetic bias field enables an improvement of the coherence time by a factor of $1.5$ to approximately $330~\mu$s. This results in a lower bound on the atom-photon entanglement fidelity of $0.952(7)$ and $0.941(7)$ (relative to a maximally entangled state) with an atomic readout delay of $26~\mu$s and $17~\mu$s in Alice's and Bob's setups, respectively. We refer the interested reader to Supplemental Material B for more details.

Finally, the quality of entangled atom-atom state is improved by optimising the two photon interference of the BSM through a rigorous analysis of the atom-photon entanglement generation process. Here, the multi-level structure of $^{87}$Rb, the finite duration of the excitation pulse, and experimental imperfections lead to the possibility of two photon emission from one atom. Crucially, these multi-photon events reduce the fidelity of the BSM result. To overcome this, only photons which are emitted after the end of the prior excitation pulse are accepted in the BSM, within a time window of $95~\text{ns}$. This reduces the entanglement generation rate by a factor of 4 (resulting in the entanglement generation rate mentioned before), but significantly increases the fidelity of the generated state; see Supplemental Material C for more details.

\subsection{DIQKD implementation} 
The independent random inputs to the devices are provided by independent quantum random number generators (QRNG)~\cite{Fuerst2010,Rosenfeld2017} located in each laboratory (requirement~iv-b). At Alice's  side, two random bits are used to select the input, while at Bob's side only one random bit is used, leading to uniformly distributed input combination choices. Based on the generated entangled state (Eq. \ref{eq:atomatom}) and the atomic state measurement scheme (Eq. \ref{eq:readout}), the input values $X\in\{0,1,2,3\}$ convert to measurement angles $\alpha \in \{-22.5^\circ,22.5^\circ,-45^\circ,0^\circ\}$ for Alice's device and $Y \in \{0,1\}$ translate to $\beta \in \{-22.5^\circ,22.5^\circ \}$ for Bob's device, respectively. The capability for fast switching between various read-out settings is achieved by overlapping multiple read-out beams with different polarisation and individually controllable intensities~\cite{Rosenfeld2017}. The outputs $A,B \in \{\uparrow,\downarrow\}$ are derived from the fluorescence counts after the state-selective ionisation. Finally, the users' inputs and outcomes are stored in two independent, trusted secure storage (requirement~iv-c). 

Unauthorised incoming and outgoing communication of the laboratories can be mitigated with prudent steps (requirements ii and iii). Especially on Bob's side, extra measures are taken to prevent information leakage from the laboratory: a free-space shutter is used during the read-out process to prevent fluorescence light from leaking out into the optical fibre (and into the outside environment) (see Fig.~\ref{fig:ExpScheme}), and the trap is always emptied before reopening the shutter. Due to the approximate $5~\text{ms}$ reaction time of the shutter, a spectral filter is deployed to block the read-out pulse after interacting with the atom and to prevent unintentional transmission of the read-out setting. For Alice's side, such countermeasures are not needed as the BSM setup already serves as a natural filter~\cite{BP2012}.

\section{Results} 
The inputs and outputs of the devices were recorded for $N=3342$ rounds over a measurement period of $75~\text{hours}$. The resulting output (anti-)correlation probabilities for the eight different input combinations, i.e. $N_{X,Y}^{A=B}/N_{X,Y}$ and $N_{X,Y}^{A\neq B}/N_{X,Y}$, are shown in Fig.~\ref{fig:result}a.

It is instructive to first review the updated performance of the QNL independently of the DIQKD protocol. Here, the figure of merit is the fidelity of the observed entangled atom-atom state relative to a maximally entangled state. By fitting the data (Fig.~\ref{fig:result}a) with sinusoidal functions, the estimated visibility for input combinations $X=2,0,3,1$ and $Y=0$ (resp. $X=2,0,3,1$ and $Y=1$) is $0.869(25)$ (resp. $0.888(45)$). Then, averaging the found visibilities and taking into account that a third atomic ground level spin state can be populated ($5^2S_{1/2}|F=1,m_F=0\rangle$), a lower bound on the fidelity is given by $\mathcal{F} \geq 0.892(19)$~\cite{Rosenfeld2017}.

The CHSH value is found to be $S=2.578(75)$ using Eq.~(\ref{eq:S}) with $E_{2,0}=-0.599(41)$, $E_{3,0}=-0.664(36)$, $E_{2,1}=0.618(39)$, and $E_{3,1}=-0.697(35)$. The QBERs are given by the correlation data for $X=Y$, i.e., $Q_0=0.0781(127)$ and $Q_1=0.0777(132)$, which gives an average error rate of $Q=0.0779(91)$. For the considered DIQKD protocol, the observed $S$ value and QBER suggest that the DIQKD setup is capable of achieving a secret key rate of $0.07$ in the asymptotic limit (Fig.~\ref{fig:result}b). To get a sense of how reliable this estimate is, we assume that underlying input-output probability distributions are independent and identically distributed and use standard Bayesian methods to determine the uncertainties of the estimated parameters. We find that taking the worst case estimates of $S$, $Q_1$, and $Q_2$ using a common probability (tail) error of 3\% give positive rates. These results indicate a proof-of-concept realisation of DIQKD on a quantum network link connecting two users $400~\text{m}$ apart. Note that the improved performance of the QNL setup even allow for the implementation of the original DIQKD protocol~\cite{Acin2007,Pironio2009}, which is more demanding than the considered protocol.

\begin{figure}
\centering
\protect\includegraphics[width=0.48\textwidth]{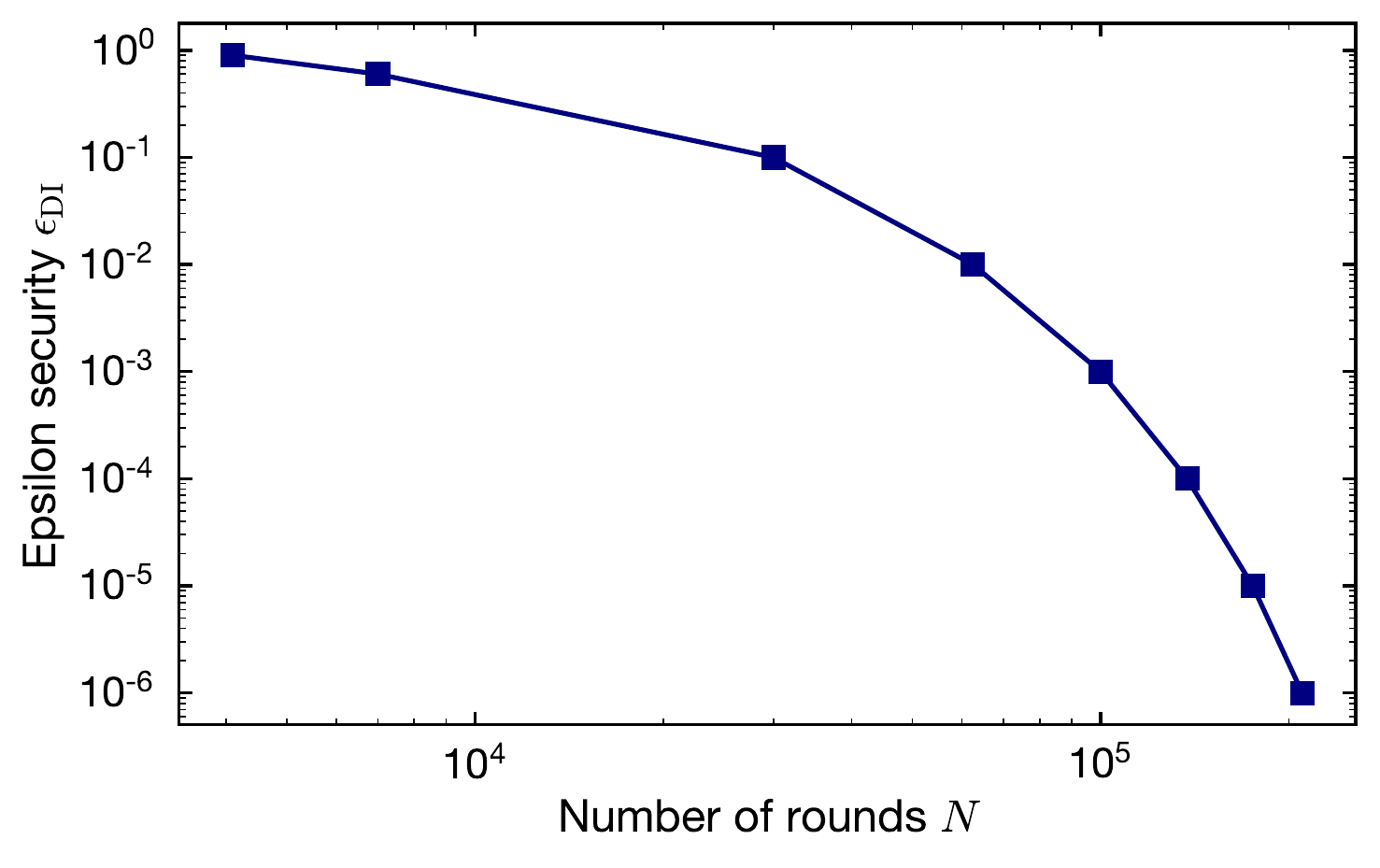}
\caption{\textbf{Finite-key simulation for the robust DIQKD protocol.} Shown is the minimum number of rounds required to distribute a finite-key with a certain epsilon security, considering collective attacks and uniformly distributed measurement settings. The channel parameters $S$, $Q_0$, and $Q_1$, are set to the observed values in the experiment. A non-asymptotic security of $\epsilon_{\textrm{DI}}=10^{-5}$ is considered to be realistic for cryptography applications.}
\label{fig:sim}
\end{figure}

Using state-of-the-art finite-key analysis for the protocol, we find that $\epsilon_{\textrm{DI}}=10^{-5}$ security can be obtained with a minimal block length of $1.75\times 10^5$~\cite{tan2020improved}, as illustrated in Fig. \ref{fig:sim}. Here, $\epsilon_{\textrm{DI}}$ is the security error of the protocol, which quantifies how close the actual output of the protocol is to the ideal output (see Ref.~\cite{tan2020improved}). In the simulation, we consider collective attacks, an error correction efficiency of 1.15, and uniformly distributed measurement settings for Alice and Bob.

\section{Outlook and Discussion}

In this work, we presented a proof-of-concept DIQKD experiment by demonstrating a QNL that could achieve positive secret key rates over 400 metres (700 metres fibre length) in a fully device-independent setting. While the current setup improves upon existing loophole-free Bell setups, there are still several areas that require enhancements before a DIQKD experiment with finite-key security and longer reach can be achieved.

For one, a significantly higher event rate is required to obtain finite-key security within a practical time-frame; based on the current setup and over the current distance, one would need months of measurement time to achieve finite-key security. 
The event rate critically depends on the entanglement generation efficiency and the repetition rate. 
To increase the former, several improvements are possible, e.g., improving the BSM setup fidelity to include the $|\Psi^-\rangle$ state projection would increase the entanglement generation rate by a factor of 2 (see Supplemental Material IV). 
However, the latter has an intrinsic limitation for event-ready schemes, such as the DIQKD scheme presented here: a repetition of the entanglement generation process is only possible after waiting for a feedback signal from the BSM. Therefore, the repetition rate is limited by the communication times between the two devices and the BSM. 
Consequently, there is a trade-off between the event rate and distance and one would have to prioritise distance over finite-key security (or vice versa).
To address this issue for the presented setup, it is possible to scale up the number of atom traps using multi-dimensional arrays~\cite{endres2016atom,barredo2016atom,de2019defect}, which combined with time multiplexing techniques~\cite{schupp2021interface} could increase the event rate by several orders of magnitude.

Another direction to improve is the reach of the QNL. Here, the limiting factor is attenuation loss of the 780 nm photons in long optical fibres, which is already 50\% for a 700 m long link. To overcome losses in longer fibre links, one can convert the entangled single photons to the low-loss telecom band via polarisation-preserving quantum frequency conversion~\cite{Leent2020, Leent2021}. Preliminary study of recent results indicates that distances up to $100~\text{km}$ are within reach.

In summary, our results represent a major step towards the goal of ultimate secure communication based solely on quantum physics. Importantly, they indicate that state-of-the-art quantum network links are capable of harnessing the ultimate quantum advantage for secure communications. Even if it is still a long way to go; when the future quantum repeater based quantum networks provide the key resource, i.e., shared entanglement, DIQKD--as realized in this proof-of-concept experiment--will become the standard for secure key exchange.

\textit{Note added in proof.} While completing the manuscript, we became aware of a similar proof-of-concept DIQKD experiment~\cite{Nadlinger2021}.

\section{Acknowledgements}
We thank Ignatius William Primaatmaja, Ernest Y.-Z. Tan, and Koon Tong Goh for useful inputs and discussions. W.Z., T.L., K.R., R.G., F.F., S.E., and H.W. acknowledge funding by the German Federal Ministry of Education and Research (Bundesministerium f{\"u}r Bildung und Forschung (BMBF)) within the project Q.Link.X (Contracts No. 16KIS0127, 16KIS0123, 16KIS0864, and 16KIS0880), the Deutsche Forschungsgemeinschaft (DFG, German Research Foundation) under Germany’s Excellence Strategy – EXC-2111 – 390814868, and the Alexander von Humboldt foundation. C.C.-W.L and R.S. are funded by the National Research Foundation, Singapore, under its NRF Fellowship programme (NRFF11-2019-0001) and NRF Quantum Engineering Programme 1.0 (QEP-P2). V.S. and C.C.-W.L acknowledge support from the National Research Foundation and the Ministry of Education, Singapore, under the Research Centres of Excellence programme.

\bibliography{main_diqkd}

\end{document}

\title{Supplemental Material for: \\ ``Experimental device-independent quantum key distribution between distant users''}

\author{Wei Zhang}
\altaffiliation{These authors contributed equally}
\affiliation{Fakult{\"a}t f{\"u}r Physik, Ludwig-Maximilians-Universit{\"a}t, M{\"u}nchen, Germany}
\affiliation{Munich Center for Quantum Science and Technology (MCQST), M{\"u}nchen, Germany}

\author{Tim van Leent}
\altaffiliation{These authors contributed equally}
\affiliation{Fakult{\"a}t f{\"u}r Physik, Ludwig-Maximilians-Universit{\"a}t, M{\"u}nchen, Germany}
\affiliation{Munich Center for Quantum Science and Technology (MCQST), M{\"u}nchen, Germany}

\author{Kai Redeker}
\altaffiliation{These authors contributed equally}
\affiliation{Fakult{\"a}t f{\"u}r Physik, Ludwig-Maximilians-Universit{\"a}t, M{\"u}nchen, Germany}
\affiliation{Munich Center for Quantum Science and Technology (MCQST), M{\"u}nchen, Germany}

\author{Robert Garthoff}
\altaffiliation{These authors contributed equally}
\affiliation{Fakult{\"a}t f{\"u}r Physik, Ludwig-Maximilians-Universit{\"a}t, M{\"u}nchen, Germany}
\affiliation{Munich Center for Quantum Science and Technology (MCQST), M{\"u}nchen, Germany}

\author{Ren{\'e} Schwonnek}
\affiliation{Department of Electrical \& Computer Engineering, National University of Singapore, Singapore}
\affiliation{Naturwissenschaftlich-Technische Fakult{\"a}t, Universit{\"a}t Siegen, Germany}

\author{Florian Fertig}
\affiliation{Fakult{\"a}t f{\"u}r Physik, Ludwig-Maximilians-Universit{\"a}t, M{\"u}nchen, Germany}
\affiliation{Munich Center for Quantum Science and Technology (MCQST), M{\"u}nchen, Germany}

\author{Sebastian Eppelt}
\affiliation{Fakult{\"a}t f{\"u}r Physik, Ludwig-Maximilians-Universit{\"a}t, M{\"u}nchen, Germany}
\affiliation{Munich Center for Quantum Science and Technology (MCQST), M{\"u}nchen, Germany}

\author{Valerio Scarani}
\affiliation{Centre for Quantum Technologies, National University of Singapore, Singapore}
\affiliation{Department of Physics, National University of Singapore, Singapore}

\author{Charles C.-W. Lim}
\altaffiliation{charles.lim@nus.edu.sg}
\affiliation{Department of Electrical \& Computer Engineering, National University of Singapore, Singapore}
\affiliation{Centre for Quantum Technologies, National University of Singapore, Singapore}

\author{Harald Weinfurter}
\altaffiliation{h.w@lmu.de}
\affiliation{Fakult{\"a}t f{\"u}r Physik, Ludwig-Maximilians-Universit{\"a}t, M{\"u}nchen, Germany}
\affiliation{Munich Center for Quantum Science and Technology (MCQST), M{\"u}nchen, Germany}
\affiliation{Max-Planck Institut f{\"u}r Quantenoptik, Garching, Germany}

\date{\today}

\maketitle

\appendix
\counterwithin{figure}{section}

\section{In which sense DIQKD is ``device-independent'': assumptions and requirements}\label{app:ass-req}

The name \textit{device-independent} QKD suggests that secrecy can be guaranteed ``without any knowledge of the device''. Such a compact statement may lead (and has actually led) to misinterpretations. It requires qualification, which we split in four requirements already mentioned in the main text. The qualified claim of DIQKD is: given devices whose inputs, outputs and interfaces are controlled by the users [requirements (i) and (ii)], secrecy is guaranteed under the obvious assumption that the secret does not leak out of the secure locations (iii), as well as under the requirements needed for any QKD protocol (iv). In this Appendix we elaborate on these matters.

\paragraph{Scenario --} Two parties, named Alice and Bob, want to establish a secret key in order to exchange secret messages. A \textit{secret key} is a list of bits that is identical between Alice and Bob, and guaranteed to known only to them -- in other words, it is \textit{shared secret randomness}. The adversary, who may be actually trying to break the protocol, is called Eve for narrative convenience. Quantum key distribution (QKD) is a practical solution for this task. The resources required in QKD are:
\begin{itemize}
    \item Secure locations: it must be assumed that the two environments, in which Alice and Bob operate, are not compromised. However, practically speaking, this can never be guaranteed unconditionally; the level of paranoia is subjective, for this involves individuals and methods which go beyond what quantum physics can certify. As such, one can only enforce the best possible known methods in practice to prevent unauthorised information leakage. For example, in our experiment, we used a free-space shutter and spectral filters to prevent fluorescence light from leaking out into the outside optical fiber.
    \item An unlimited (for all practical purposes) amount of local randomness, i.e.~the possibility of generating strings of bits that are unknown to anyone else (in this case, even to the other authorised partner). These will constitute the ``trusted inputs'' to the devices.
    \item An authenticated public channel for classical communication between them. In order to authenticate the channel, Alice and Bob must possess some shared randomness prior to the start. Thus, QKD is actually quantum key \textit{expansion}: the amount of secret key generated by the protocol should exceed the amount that is consumed to authenticate the channel and for classical post-processing.
    \item Last but not least, the actual devices that create and process the quantum information, and the quantum channel connecting them. \textit{The ``device-independence'' of DIQKD means that these devices can be dealt with as black boxes}. Explicitly, the security assessment does not rely on the characterisation and modeling of any of their inner workings and dimensions, not even the type of quantum system and measurements that are actually performed.
\end{itemize}

Any QKD protocol essentially starts with the distribution and measurement of quantum signals. This part consists of well defined rounds, whereby each round consists of one pair of inputs and outputs for each device. After accumulating a certain amount of rounds the device inputs are shared over the trusted public channel. Certain input combinations are then used to generate the \textit{sifted (or raw) key}, while others are used to estimate the features of quantum mechanics used to bound Eve's information on the outputs. It is then possible to proceed with error correction and privacy amplification protocols, and extract a \textit{final key} on which Eve has no information. These steps also require adequate and trusted methods that fit to the actual implementation and performance of the DIQKD setup, in order to not distort the information theoretical security.

For DIQKD, the feature of quantum theory used to bound Eve's information is \textit{the violation of a Bell inequality}~\cite{Acin2006,Acin2007,Pironio2009,Vazirani2014,Arnon-Friedman2018}. A Bell inequality test has its own set of requirements, failure to comply with which leads to famous loopholes~\cite{Larsson2014,Scarani-book}. First, the requirement of locality, ensuring that the process generating the output in one device is independent of the input and the process of the other. Second, in each round the inputs should be random for the devices. Notice that this is slightly different from the analog requirement of QKD: for QKD, the local input should be random for Eve but might be known by the device; for Bell alone, the local input may be publicly known, as long as it is random for the device. In DIQKD, the local randomness should therefore be \textit{random both for Eve and for the device}. Among other loopholes that may invalidate a Bell test, by far the most important and relevant here is the ``detection loophole''. To avoid it, one must not assume fair sampling in case of imperfect detection efficiencies: rather, there must be an output of the device for every input (if the detector failed to detect, the output must be generated according to some other local recipe: this will of course reduce the correlations, but won't compromise the soundness of the test). In fact, it is the detection loophole opened by losses that makes it very challenging to implement DIQKD with purely optical setups.

Finally, even though the devices are black boxes, for secrecy one should require that they do not leak any information. For one, the provider of the devices should be trusted as honest: if they are colluding with Eve, surely they have hidden somewhere a small emitter that might leak the key at the end of the protocol (or in later instances). On a more technical level, these devices must be open to the world through the quantum channel (the quantum signals, however uncharacterised, must be able to enter the device). One must then assume that no information leaks out through that port, while open~\cite{Arnon-Friedman2019}. Once again, the assumption of no-leakage from the secure location is a requirement for all forms of secrecy. We just brought up two possible forms of leakage that are worth mentioning, given the danger of exaggerations associated with the words ``device-independence''.

Based on what we said, we can summarize the requirements for DIQKD in the following four (order does not indicate importance):
\begin{itemize}
 \item[(i)] The used system consists of two separated devices, the devices receive an input and respond with a well defined output, and the protocol is split into well defined rounds;
 \item[(ii)] Alice and Bob control when the devices communicate with each other;
 \item[(iii)] The devices do not send classical information to an possible eavesdropper;
 \item[(iv-a)] Quantum theory is correct;
 \item[(iv-b)] Each device is supplied with trusted inputs independent and unknown to an possible attacker (Eve);
 \item[(iv-c)] Alice and Bob are connected via an authenticated channel,  employ trusted local storage units, and use appropriate post processing.
\end{itemize}

\paragraph{Experimental requirements--}
Each of the listed premises has different consequences for an implementation of DIQKD. (iv-a) is obvious for any kind QKD. It means that if the world is described by a more advanced theory than quantum mechanics the security proof might not hold. This has, however, no consequences for implementations. The other premises can be placed in two categories. The first, containing only (i), leads to requirements for the devices which need to be addressed by the manufacturer. The second category (ii), (iii), (iv-b), and (iv-c) leads to requirements on the operational environment of the devices, which need to be addressed by the users, Alice and Bob. 

(i) seems obvious and is most often not stated explicitly as an premise but only mentioned indirectly when describing the protocol. Moreover, it is important to note that DIQKD is possible with any number of devices that is above two. But there needs to be at least one device for each party. One large device that connects both labs contradicts the assumptions of a Bell test and renders compiling to premises (ii) and (iii) impossible. Further, the devices must give an unambiguous output when provided with an input and should be easy to identify even for non-experts. Defining a microscopic quantum object, e.g., an atom  as an device is in principle possible for DIQKD in contrast to other device-independent applications (random number generation~\cite{Pironio2010} or self-testing~\cite{bancal2021self}). However, such a definition is not very useful since a quantum object will always be embedded in a bigger device holding and controlling it and hence this can be defined as the device without bothering about the actual quantum system. The well defined rounds are necessary as the Bell test demands that for each input one always receives an output, otherwise the detection loophole will be open and invalidate the DI trust. Therefore, (i) directly transforms for a requirement for a system designed for DIQKD.

Now to the requirements that need to be addressed by the users, Alice and Bob. Restricting the communication of the devices for DIQKD is necessary, as it  (ii) ensures local measurements for the Bell test and (iii) prohibits the possible malicious devices from simply leaking information to an eavesdropper. In many works, including~\cite{Schwonnek2020}, these two premises are combined to the demand for perfectly shielded rooms for Alice and Bob. However, such an ideal room is in practice impossible to realize without at least assuming some limitations on the devices. The biggest obstacle is that the two devices need to establish entanglement between them. This can be realized in different ways, but in all of them there is some physical connection to the outside-world. Prohibiting information leakage over this connection cannot be guaranteed without additional assumptions. Nevertheless, it is possible to build very secure rooms to limit the possibility of information leakage dramatically, dependent on the demands of the user.

Furthermore, trusted inputs (iv-b) are necessary for QKD as well as for a Bell test. They are best provided by trusted random number generators, which are indeed a common demand for cryptographic application. However, here is not the explicit need of a true or quantum random number generator, one can also use any bit sequence which is unknown to the devices and potential eavesdroppers.

Finally, the last premise (iv-c) summarizes all necessities for the extraction of a secret key from the recorded data. These are not always explicitly stated but are then implicitly still made. The authenticated classical public channel is needed to ensure that the DIQKD connection is between Alice and Bob and not relayed to an eavesdropper for, e.g., man-in-the-middle attacks. The trusted storage is needed to ensure the integrity of the recorded data. Storing in- and outputs only in the devices is not possible as they might be malicious and simply exchange the recorded inputs and outputs with a prerecorded data set. Although not discussed in detail here, the appropriate methods for error correction and privacy amplifications have to be used. 

\paragraph{Proof-of-concept implementation}

The first step from the proposal to a real world application is a proof-of-concept implementation. Here, the goal is to show that the protocol can be implemented with the currently available technology. For such an experimental implementation some of the requirements can be relaxed, as the goal is not to send a secret message, but to show that this is in principle possible. This is especially true for the requirements that need to be addressed by the users. Thus, the main goal is to build two devices that fulfill requirement (i) and permit users to fulfill requirements  (ii), (iii), (iv-b), and (iv-c).

As described in the main text, the presented QNL formed by the two atom traps enables exactly this. It consist of two independent devices. The devices are able to receive four, respectively two, different input values and respond with an unambiguous output. The heralded entanglement generation and event-ready measurement scheme allow for well defined rounds and closes the detection loophole in a Bell test. Thus it is compatible with all demands in (i). To further prove it is compatible with the other requirements they are fulfilled in a reasonable fashion, see the main text. This shows the suitability of the proof-of-concept implementation without the need of further argumentation, e.g., based on the physical model of the devices.

\section{Atom-Photon Entanglement Generation}
\label{app:atom-photon}

Both devices, i.e. atom traps, are characterized individually by analyzing the atom-photon entanglement generation process. The process starts by preparing the atom in the $5^{2}S_{1/2}|F=1, m_F=0\rangle$ state, denoted as $|1,0\rangle$, via optical pumping. Next, the atom is excited with a laser pulse that is resonant to the transition $5^{2}S_{1/2}|F=1\rangle\rightarrow5^{2}P_{3/2}|F'=0\rangle$ and polarized parallel to the quantization axis ($\pi$-polarization). The temporal shape of the pulse is approximately Gaussian ($22~\text{ns}$ FWHM). In the subsequent decay, the polarization of the photon that is emitted along the quantization (z-)axis becomes entangled with the atomic spin state, resulting in the following maximally entangled atom-photon state

\begin{align*}
|\Psi\rangle_{AP} & =  1/\sqrt{2}(|\downarrow\rangle_z|L\rangle + |\uparrow\rangle_z|R\rangle) \\
 & = 1/\sqrt{2}(|\downarrow\rangle_x|V\rangle + |\uparrow\rangle_x|H\rangle),
\end{align*}

\noindent where $|\downarrow\rangle_z$ and $|\uparrow\rangle_z$ denote atomic spin states $|1,-1\rangle$ and $|1,+1\rangle$, $|L\rangle$ and $|R\rangle$ denote left- and right-circular photonic polarization states, and $|V\rangle$ and $|H\rangle$ denote vertical and horizontal linear photonic polarization states, respectively. 

The success probability of the entanglement generation process, i.e. detection of a photon after an excitation pulse, equals $5.98 \times 10^{-3}$ and $1.44 \times 10^{-3}$ for Alice's device and Bob's device, respectively. Note that the lower photon detection probability for Bob's device is due to attenuation loss of approximately $50\%$ in the $700~\text{m}$ optical fiber and the loss due to additional optical elements, including the beam splitter (90:10) for the local fluorescence detection and spectral filter shielding the read-out light, by another $50\%$, see Figure~2 of the main text.

The atomic spin state is analyzed after a delay of $25.55~\text{µs}$ and $16.7~\text{µs}$, for Trap 1 and 2, respectively. This time allows for event-ready entanglement generation (two-way communication time between the labs equals approximately $7~\text{µs}$) and provides rephasing of both the Larmor precession due to the magnetic bias field $57~\text{mG}$ and $168~\text{mG}$ along the y-axis and the transverse trap frequencies.

The atomic qubit is analyzed via a state-selective ionization scheme~\cite{Leent2020,PhD-Norbert}, see main text Fig. 3b. There, a particular state of the atomic qubit is transferred to the $5^2P_{1/2}|F'=1\rangle$ depending on the polarization $\chi=\cos(\gamma)V+e^{-i\phi}\sin(\gamma)H$ ($\gamma = \alpha$ for Alice's and $\gamma = \beta$ for Bob's device) by a $140~\text{ns}$ laser pulse from where it is ionized by a bright $473~\text{nm}$ laser pulse and thus leaves the trap. If the atom is still in the trap it is thus projected onto the state 
\begin{equation}
|\textrm{Dark}\rangle=e^{i\phi}\cos(\gamma)|\uparrow\rangle_x-\sin(\gamma)|\downarrow\rangle_x.
\end{equation}

\noindent In the experiment the presence of the atom is tested using fluorescence collection finally yielding the measurement outcome. 

\begin{figure}
\begin{centering}
\includegraphics[width=0.45\textwidth]{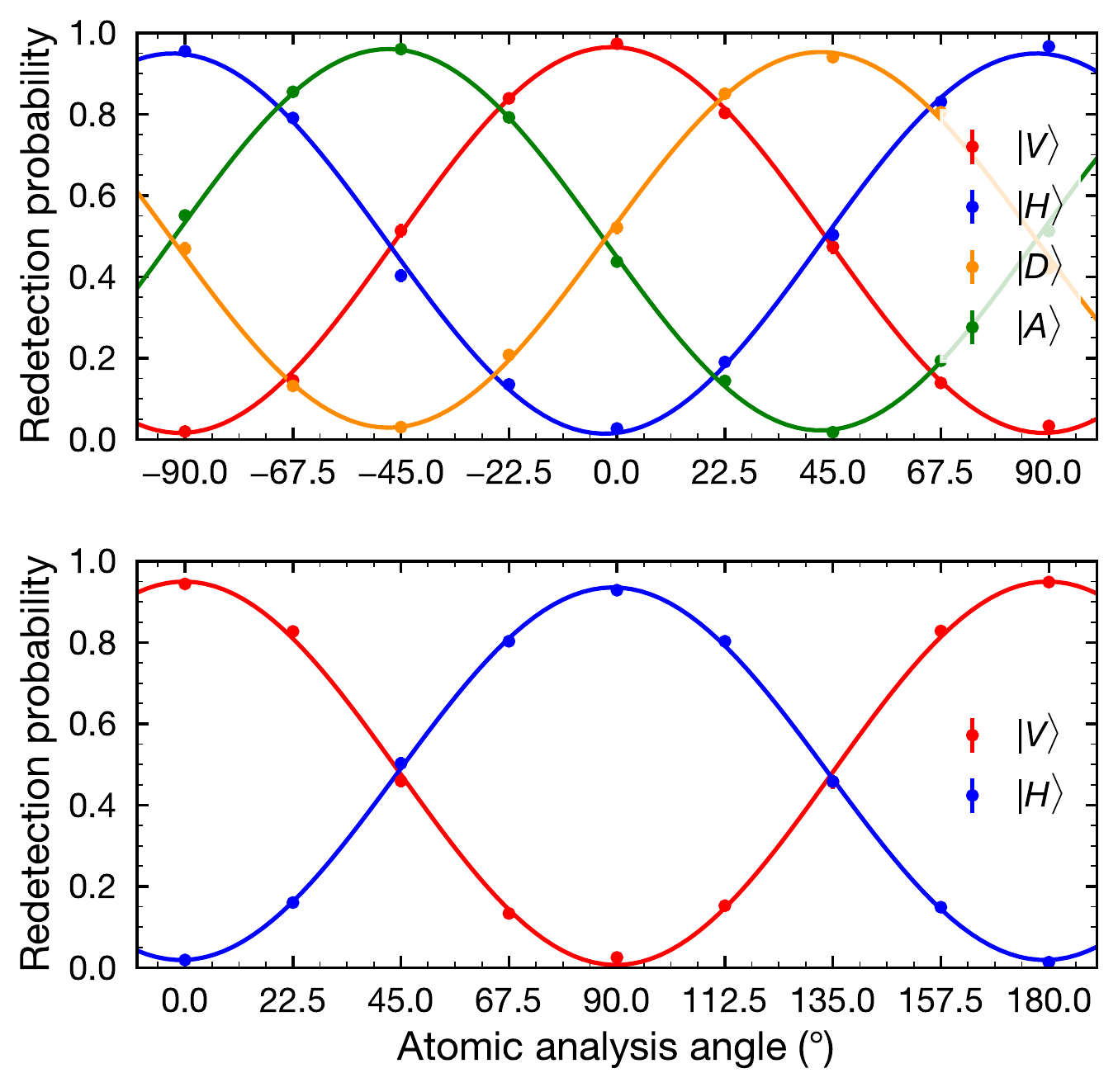}
\par\end{centering}
\caption{\textbf{Observation of atom-photon entanglement for Alice's device (top) and Bob's device (bottom).} The atomic analysis angle, i.e. readout polarization angle whereby $0\degree$ equals vertical, see equation (3) main text, is varied while measuring the photonic polarization in the $H/V$ and $D/A$ (only for Device 1) basis. Based on the fits the estimated fidelity of the entangled atom-photon state equals 0.952(7) and 0.941(7), for Alice and Bob, respectively.}
\label{fig:atom-photon}
\end{figure}

The atom-photon entanglement is analyzed by measuring the photonic polarization in the $H/V$ (horizontal/vertical) and $D/A$ (diagonal/anti-diagonal) basis, while varying the atomic analysis angle, i.e. readout polarization, as shown in Fig.~\ref{fig:atom-photon}. In test runs, we observed 35259 and 20001 events for Alice's and Bob's side, respectively. The visibilities (Vis) of the measured states are obtained by fitting the data with sinusoidal functions. These result  for Alice in visibilities of 0.942(14), 0.930(17), 0.942(13), and 0.954(19), for vertical $|V\rangle$, horizontal $|H\rangle$, diagonal $|D\rangle$, and anti-diagonal $|A\rangle$ photonic linear polarization states, respectively. For Bob, the fits give visibilities of 0.943(16) and 0.917(8), for $|V\rangle$ and $|H\rangle$ photonic linear polarization states, respectively.

To estimate a fidelity of the entangled state, one needs to take into account that a third atomic spin state can be populated $5^2S_{1/2}|F=1,m_F=0\rangle$ due to magnetic fields. Hence, assuming depolarizing noise in the 2x3 state space, a lower bound on the fidelity relative to a maximally entangled state is given by

\begin{equation}
F \geq 1/6 + 5/6 \overline{\textrm{Vis}},
\end{equation}

\noindent with the average visibility $\overline{\textrm{Vis}}$, which results in estimated fidelities of 0.952(7) and 0.941(7), for Alice's device and Bob's device, respectively.

\section{Improving the Atom-Atom Entanglement Quality}
\label{app:Time-filtering}

The quality of the entangled atom-atom state depends on the generated atom-photon entanglement in both traps (see App. \ref{app:atom-photon}) and on the performance of the Bell state measurement (BSM) on the photons. In order to understand these processes and subsequently improve on their performance, we modeled the excitation of a $^{87}\text{Rb}$ atom by a short laser pulse. Here, not only the physical properties of the system are considered, e.g., multilevel structure of the atom and the frequency broadening of the short laser pulse, but also imperfections of the experimental setup and procedure, such as imperfect polarization and state preparation. 

\begin{figure}
\begin{centering}
\includegraphics{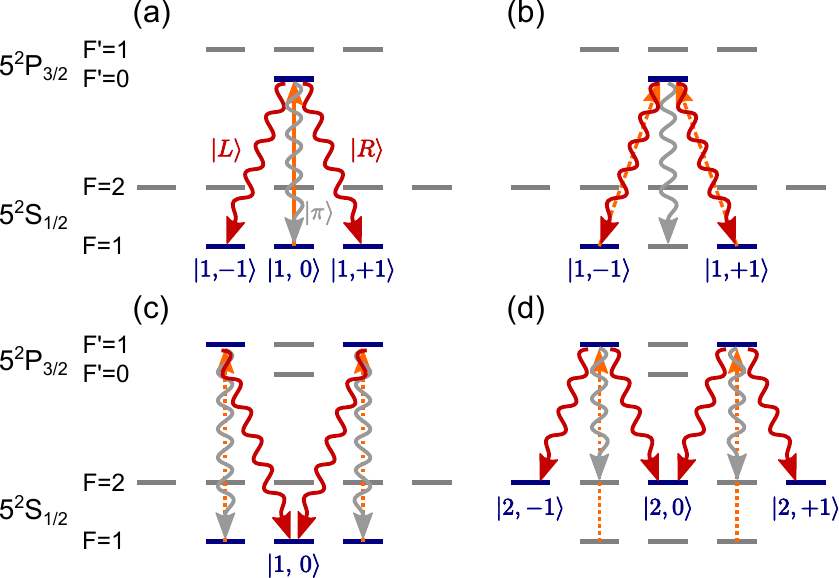}
\par\end{centering}
\caption{\textbf{Different branches of the excitation process in the level structure of $\mathrm{^{87}Rb}$.} (a) Intended generation of atom-photon entanglement in the spontaneous decay of the excited $5^{2}P_{3/2}|F'=0,m_F=0\rangle$ level. The orange arrow indicates the prior excitation pulse. Photons polarized linearly along the quantization axis ($\pi$-decays, gray) are not detected by the single photon detectors. (b) In the case of an imperfect state preparation or a first the first decay, a excitation is possible due to polarization misalignment, or (c) off-resonant excitation. (d) When the $5^{2}P_{3/2}|F'=1\rangle$
is excited, decays to $5^{2}S_{1/2}|F=2\rangle$ level is possible.
\label{fig:Levels}}
\end{figure}

In the intended atom-photon entanglement generation process (Fig.~\ref{fig:Levels}(a)) the selection rules prohibit a second interaction with the $\pi$-polarized excitation laser. However, there are two effects that result in different types of emission. The first is caused by an experimental limitation: a small polarization misalignment of the excitation laser makes a second excitation possible, see Fig.~\ref{fig:Levels}(b). Secondly, due to the small separation of the  $5^{2}P_{3/2}|F'=0\rangle$ to the  $5^{2}P_{3/2}|F'=1\rangle$ level off-resonant scattering via this level is possible (Fig.~\ref{fig:Levels}(c),(d)). These effects lead to the emission of a second photon that perpetuates the atom-photon state and reduce its fidelity. Accordingly, this will be passed through by the swapping process also reducing  the atom-atom state fidelity.

Beyond the effects reducing the fidelity of both the atom-photon and atom-atom states, there is a unapparent other effect reducing the fidelity of the two-photon interference based BSM. This process includes emission of a $\pi$-polarized photon followed by a regular excitation and decay. The $\pi$-polarized photon is not coupled into the single mode fiber and thus does not contribute to the atom-photon state, however, the temporal shape of the collected (second) photon is different than for a photon originating from a single excitation and emission process. This reduces the two-photon interference contrast, leading to a lower BSM fidelity and imperfect atom-atom state preparation.

A numerical simulation of the temporal behavior yields a time dependent photon emission (and thus detection) probability broken down for each of the different excitation processes described in Figure~\ref{fig:Levels}. A complete and detailed description of the model used for the numerical smulation can be found in~\cite{PhD-Julian,PhD-Kai}.  Based on this result it is possible to optimize the two-photon acceptance time window for the BSM. The main finding is that the resulting entangled atom-atom state has the highest fidelity relative to the desired Bell state if only photons are accepted that are emitted after the end of the excitation pulse. This excludes the perpetuated atom-photon states as well as the effect of the imperfect state preparation, and increases the quality of the entanglement swapping operation. 

While the first point follows directly from the simulated time dependent detection probability of the different excitation branches, the second is not that obvious. For this the following situation has to be considered: One atom emits only one photon, which is collected and detected, while the other one of the two atoms undergoes a two photon emission process first emitting a $\pi$-polarized photon and then being excited again emitting a second photon which is detected. If in this case one of the two detected photons is detected at an earlier time, especially during the excitation pulse, it can be assigned with very high probability to the atom emitting only one photon and the late photon to the atom with the two photon emission. Since the emission of the first $\pi$-polarized photon, in principle, allows the identification of the atom with the two photon process, the atom-atom state is not projected onto an entangled state by the BSM. 

\begin{figure}
\begin{centering}
\includegraphics[width=0.45\textwidth]{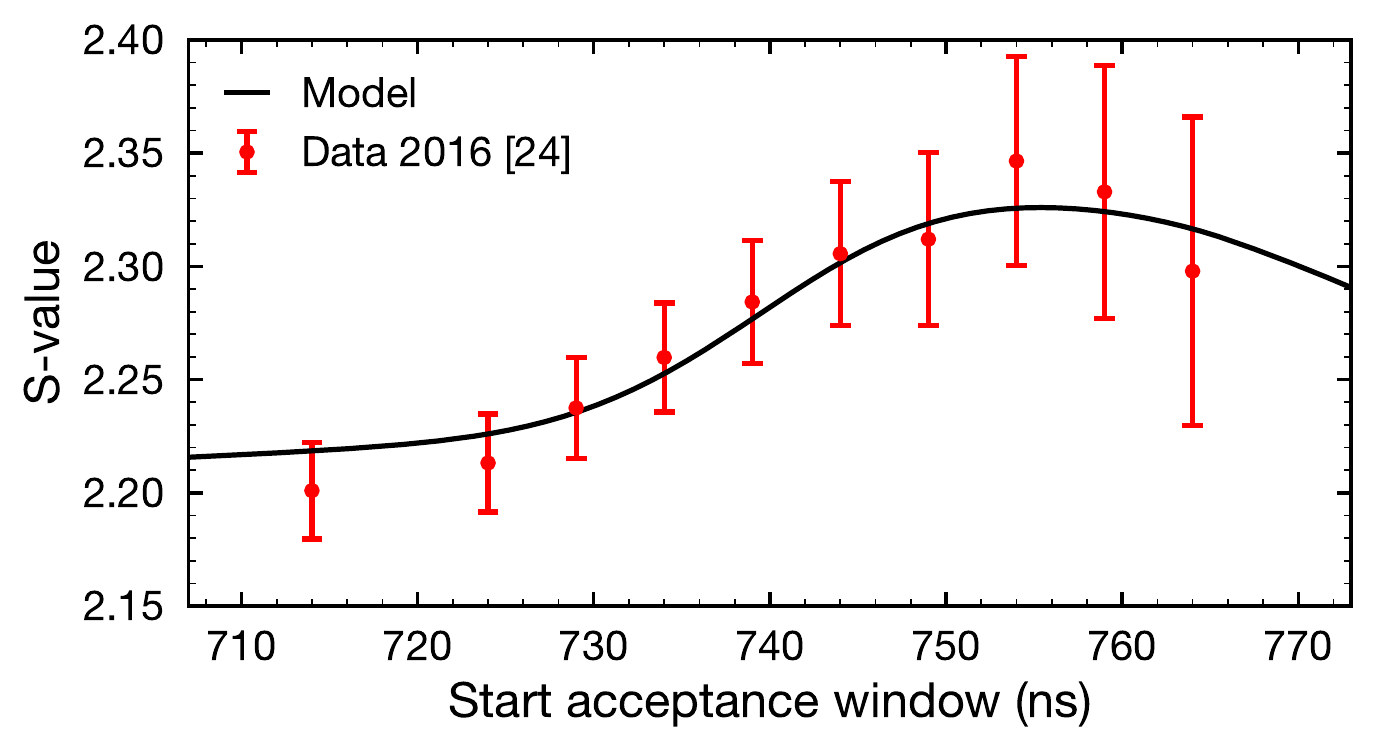}
\par\end{centering}
\caption{\textbf{CHSH S-value and relative event rate as a function of the starting time $t_s$ of the acceptance time window.} The acceptance window ends with $t_e=850~\text{ns}$ . The numerical model is compared to experimental data collected for~\cite{Rosenfeld2017}.}
\label{fig:filtering-model}
\end{figure}

\begin{figure}
\begin{centering}
\includegraphics[width=0.45\textwidth]{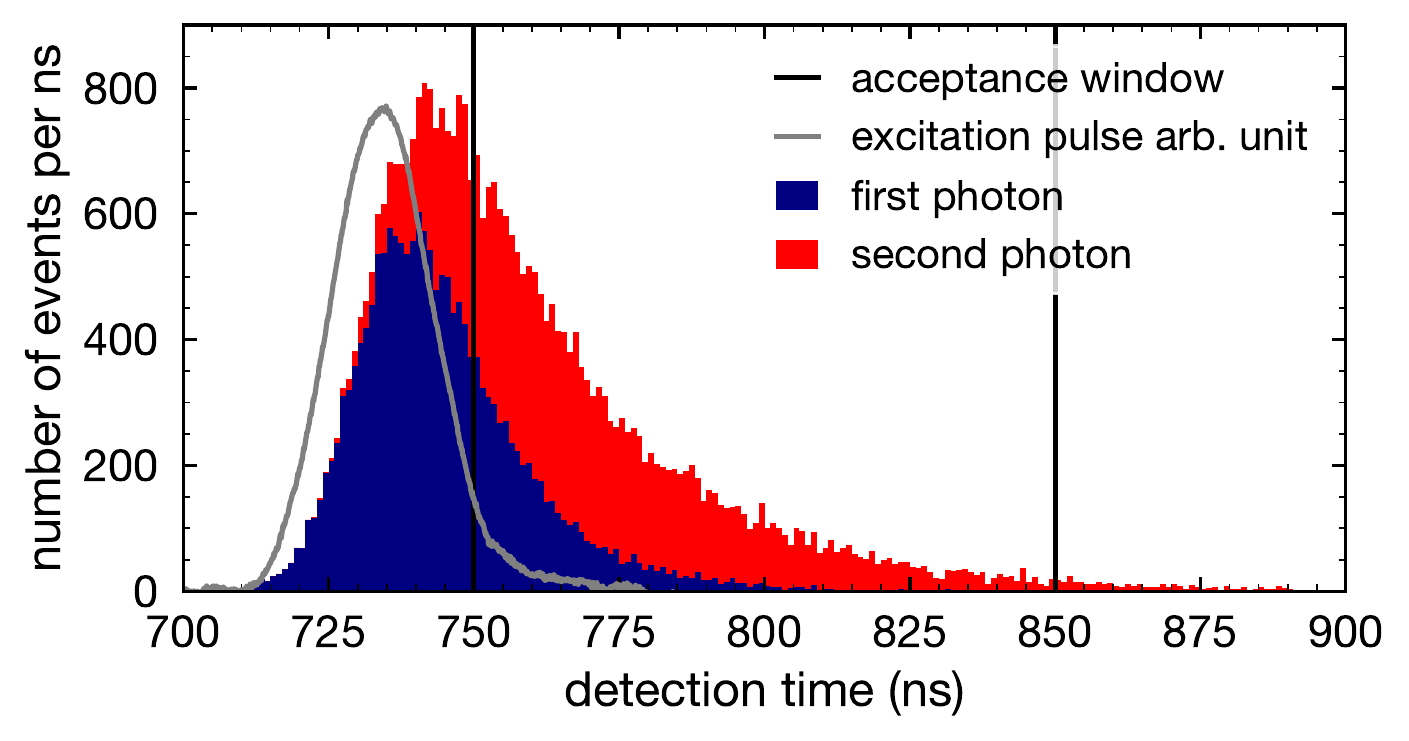}
\par\end{centering}
\caption{\textbf{Photon detection time histogram and acceptance window} In total $27\%$ of the two-photon events are accepted, note that both photons should arrive within the acceptance window. The time on the x-axis in the plot is in relation to an trigger signal provided by the control electronics of the the experiment. The position of the excitation pulse compared to the detected photons represents the timing during the emission and not the detection process.}
\label{fig:photon-shape}
\end{figure}

Based on the outcome of the model (Fig.~\ref{fig:filtering-model}), we define a two-photon acceptance time window of $95~\text{ns}$ that starts after the excitation pulse, as illustrated in Fig.~\ref{fig:photon-shape}. While it drastically increases the entanglement fidelity, as shown in the simulation and the data presented in the main text, the shorter acceptance time window reduces the event rate by a factor of $4$. Note that defining a smaller acceptance time window before the experiment does not lead to a ready-signal in the first place and thus does not open any kind of loophole, e.g., the detection loophole, in an Bell test.

\begin{figure}
\begin{centering}
\includegraphics[width=0.45\textwidth]{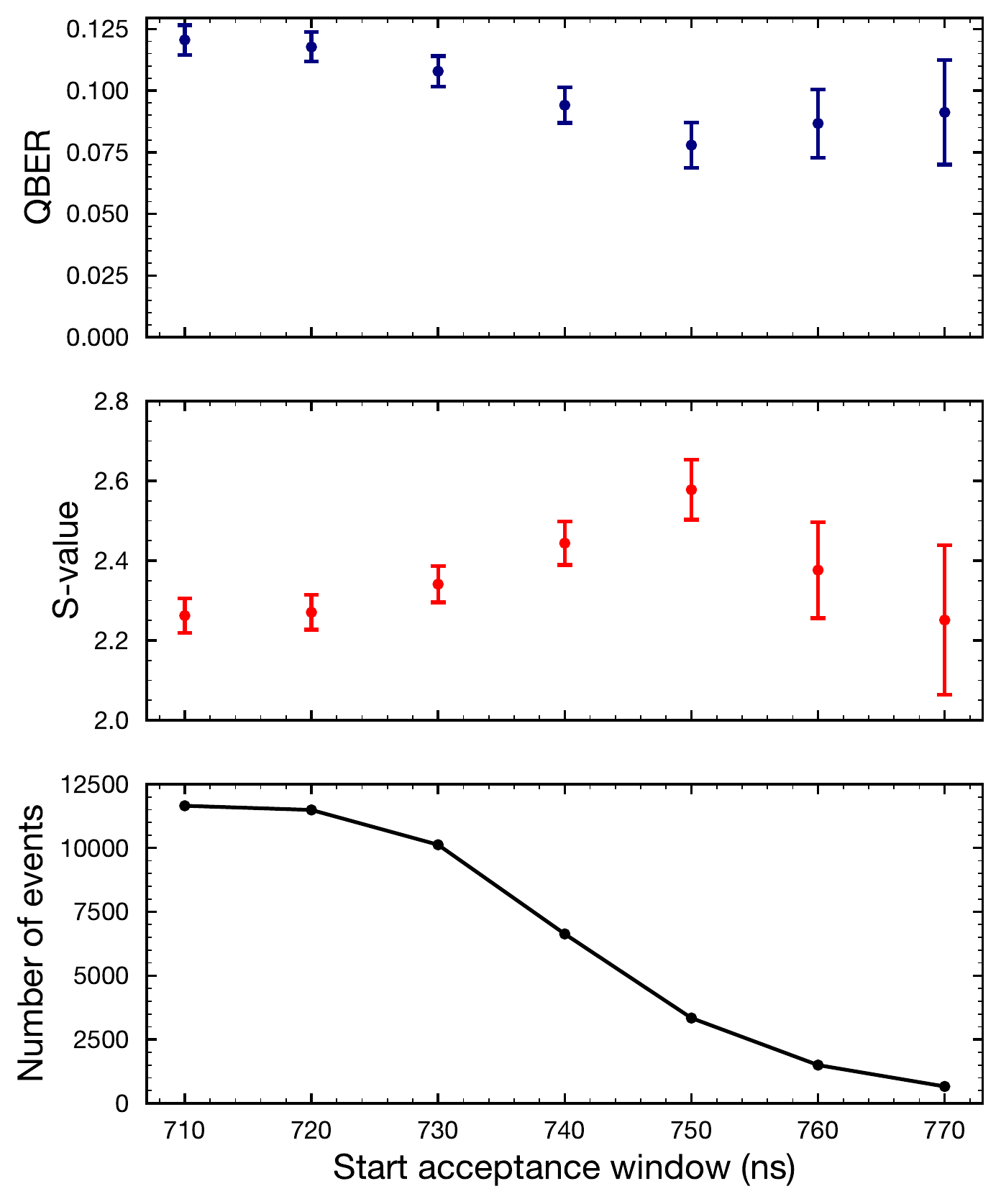}
\par\end{centering}
\caption{\textbf{Analysis of the data recorded during the DIQKD experiment.} QBER, CHSH S value, and number of events depending on the starting time of the acceptance time window. }
\label{fig:filtering-S-and-Q}
\end{figure}

\begin{figure}
\begin{centering}
\includegraphics[width=0.45\textwidth]{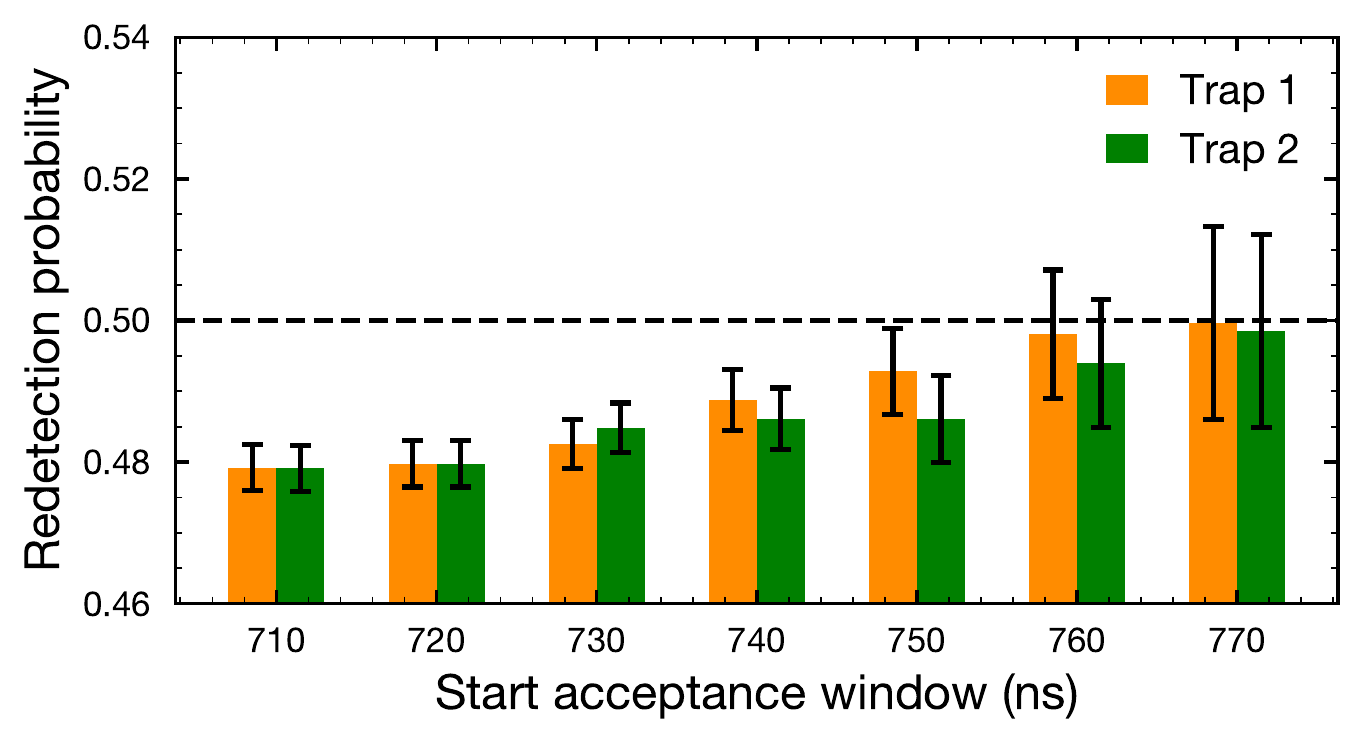}
\par\end{centering}
\caption{\textbf{Atomic state readout result for Trap 1 and 2 for varying start times of the acceptance window.} Perfect entanglement generation and readout would result in a redetection probability of 0.5. The data presented in the main text. }
\label{fig:ion-prob}
\end{figure}

For a complete analysis of the experiment, we also recorded the events outside of the time window, however, these events are not used in the DIQKD demonstration. The analysis of the complete dataset shows an increase of S-value and a reduction of the QBER for smaller time windows (Fig.~\ref{fig:filtering-S-and-Q}).  The effect of excluding events with errors in the atom-photon entanglement generation is also observed in the read-out outcomes for both traps individually, as illustrated in Fig.~\ref{fig:ion-prob}. For an ideally entangled atom-photon state, the ionization probability is $0.5$, however, the processes reducing the atom-photon state fidelity, e.g., a second off-resonant excitation, lead to atomic states with higher ionization probabilities.

\begin{figure}
\begin{centering}
\includegraphics[width=0.45\textwidth]{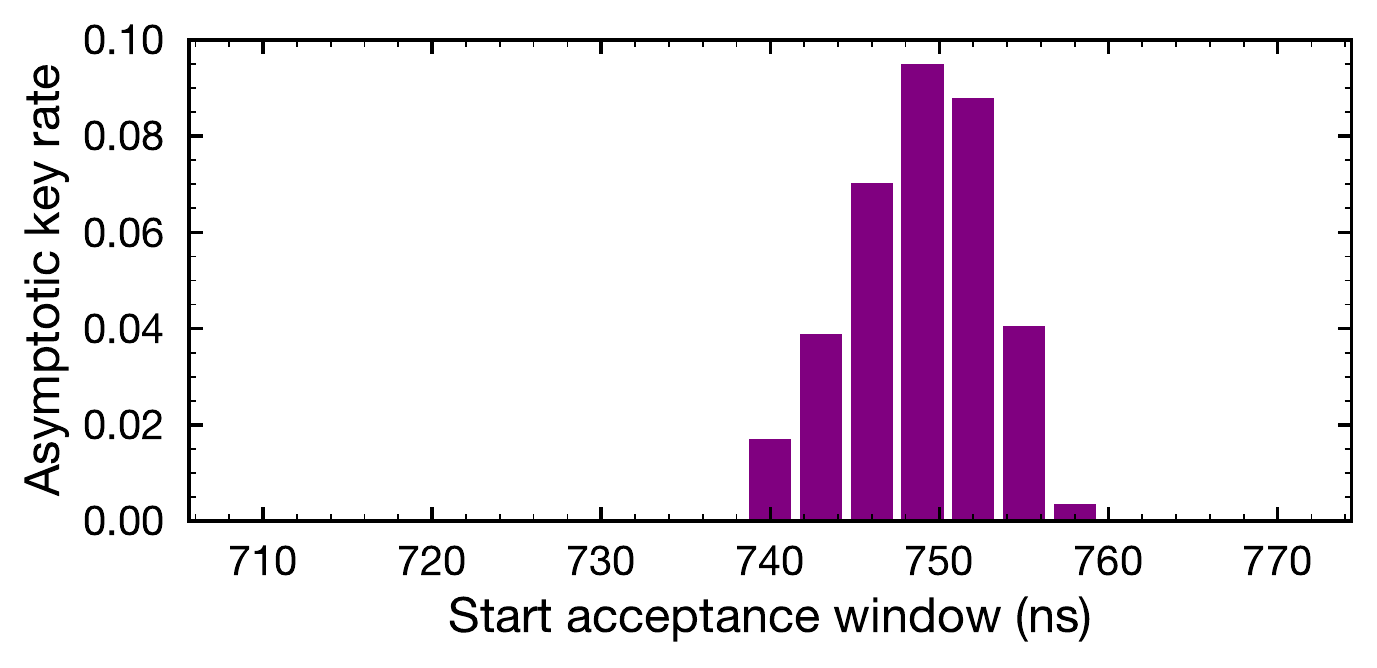}
\par\end{centering}
\caption{\textbf{Expected secret key rate for the robust DIQKD protocol for different $t_s$.}}
\label{fig:key-rate}
\end{figure}

An even smaller time window might increase the atom-atom entanglement generation even further, thus leading to higher S and lower QBER which in turn result in an higher asymptotic key rate (Fig.~\ref{fig:key-rate}). However, this further reduces the event rate and increases the time needed for a measurement yielding a sufficient amount of events. More interesting for future experiments is the possibility of optimizing the excitation pulse shape in combination with narrow band filtering of the single photon frequency. A shorter excitation pulse, in combination with spectral filtering of off-resonant excitations, might lead to a more precise filtering of unwanted photons and an higher event rate. 

To reach event rates high enough to generate a secure key using DIQKD for a finite block length~\cite{Arnon-Friedman2018}, one has not only to consider the quality of the entanglement but also the generation rate. Thus, for finding the optimal acceptance time window for such an experiment must consider the trade-off between them.

\section{Estimating the expected secret key rate}\label{app:statistics}

A rigorous security analysis of practical DIQKD would require a finite-key analysis that takes into account the resources consumed and block-length considerations~\cite{tan2020improved}. However, as mentioned in the main text, our experiment, which prioritises the establishment of swapped entangled trapped atoms 400 metres apart, has an intrinsic limitation on the event rate based on state-of-the-art technology. Consequently, there is a trade-off between the event-rate and separation of the laboratories, hence it is not realistic to demonstrate finite-key security based on known calculation method~\cite{tan2020improved}.

To that end, we estimate the expected secret key rate of the DIQKD experiment using standard Bayesian analysis; while we acknowledge that this is not the usual approach for QKD, it nevertheless gives a reliable estimate based on available data. Starting from the data summary listed in Tab. \ref{tab:data}, we model the random behaviour of $S$ (its winning probability), $Q_0$, and $Q_1$ using Beta random variables, $\beta_{\rm{win}}$, $\beta_{Q_0}$, and $\beta_{Q_1}$, respectively, which is in line with the self-testing statistical analysis reported in Ref. \cite{bancal2021self}. In particular, using a uniform prior, the (updated) posterior distributions are
\begin{table}[]
    \centering
    \caption{\textbf{Summary of in- and output measurement correlations.} Number of rounds for each of the eight input setting combinations together with the number of rounds the devices gave correlated outcomes.}
    \begin{tabular}{c c c c c}
    \hline
        \multicolumn{5}{c}{number of rounds $N_{X,Y}$} \\ \hline
             & X=0     & X=1     & X=2     & X=3 \\
         Y=0                 &  448      &    425       &   389        & 434\\
         Y=1                 &  408      &    412       &   403        & 423\\ \hline
        \multicolumn{5}{c}{with correlated outputs $N^{A=B}_{X,Y}$} \\ \hline
             & X = 0     & X = 1     & X = 2     & X = 3 \\
        Y = 0                 &  35      &    205       &   78        & 73\\
        Y = 1                 &  198      &    32       &   326        & 64\\ 
        \hline
        \multicolumn{5}{c}{with uncorrelated outputs $N^{A \neq B}_{X,Y}$} \\ \hline
             & X = 0     & X = 1     & X = 2     & X = 3 \\
        Y = 0                 &  413      &    220       &   311        & 361\\
        Y = 1                 &  210      &    380       &   77        & 359\\ 
        \hline
    \end{tabular}
    \label{tab:data}
\end{table}
%In detail, we model   
% \begin{align}
% F_S&=8\mathrm{B}(a=N_S \cdot S_{win} - 1,b=N_S(1-S_{win}) + 2)-4\nonumber \\
% F_Q&=\mathrm{B}_Q(a=N_Q \cdot Q + 2,b=N_Q(1-Q) -1),
% \end{align}
\begin{align}
\beta_{\rm{win}}&=\mathrm{Beta}(1355+1 ,1649-1355+1),\nonumber \\
\beta_{Q_0}&=\mathrm{Beta}(35+1 ,448-35+1),\nonumber \\
\beta_{Q_1}&=\mathrm{Beta}(32+1 ,412-32+1),
\end{align}
where $\mathrm{Beta}(a,b)$ is the standard Beta distribution, and the winning probability is related to the CHSH value by $P_{\rm{win}}=(S+4)/8$; thus $\lfloor 1649\times (2.578+4)/8\rfloor=1355$.

Then, to calculate the worst-case estimate of the expected secret key rate, we fix the tail errors of the updated Beta distributions to 3\%; this means a 97\% chance that each of the parameters would be higher (or lower) than a certain critical threshold. More specifically, we find $S \geq 2.4256$ and $Q_0=Q_1\leq 0.107$. Finally, using uniform settings (as was done in our experiment), we find that these critical values provide positive key rates. 

\section{Towards DIQKD Applications}\label{app:atom-arrays}

For a practical demonstration of DIQKD, the employed apparatus should:

\begin{enumerate}
\item show entanglement quality enabling for a positive key rate;
\item obtain entanglement over distances relevant for cryptography; and
\item reach entanglement rates allowing for key distribution on practical timescales.
\end{enumerate}

\noindent Currently, generating an high-quality atom-atom entanglement event per approximately $80~\text{s}$ over a distance of $400~\text{m}$, the setup does not yet fulfil the third requirement. Hence, next steps involve improving the entanglement generation rate. 

Three realistic improvements on the current setup can increase the event rate by an order of magnitude. First, preparing more than one entangled state in the swapping process, here the $|\Psi^-\rangle$ state. This is already possible with the current setup, but its quality needs to be improved. Second, using superconducting nanowire single-photon detectors with quantum detection efficiencies of $>90\%$, which could quadruple the entanglement generation rate. Finally, improving the atom-photon entanglement generation quality and hence reducing the requirement of temporal filtering in the BSM, see Appendix \ref{app:Time-filtering}.

Beyond these incremental improvements, neutral optically trapped atoms are an ideal candidate to scale up the number of individually controllable atom traps and hence enable for temporal multiplexing of the entanglement generation process. By employing micrometer spaced trapping potentials, it is possible to realize defect free arrays of single atoms while allowing for individual control of the trapping sites.

Various approaches exist to realize multi-dimensional trap arrays, for example, using a spatial light modulators~\cite{barredo2016atom}, acousto-optical deflectors ~\cite{endres2016atom}, or microlens array~\cite{de2019defect}. Currently, state-of-the-art trapping techniques allow for individual storage and control of $>100$ single atoms, potentially increasing the event rate by orders of magnitude.

Another advantage of employing arrays of single atom traps is the possibility to implement entanglement distillation protocols when sharing various entangled atom-atom pairs between two setups~\cite{kalb2017entanglement}. This provides an promising platform to realize a quantum repeater.

\bibliography{supp_diqkd}